\documentclass[%
 aip,
 amsmath,amssymb,
 reprint,%
]{revtex4-1}

\usepackage{graphicx}
\usepackage{dcolumn}
\usepackage{bm}

\usepackage[utf8]{inputenc}
\usepackage[T1]{fontenc}
\usepackage{mathptmx}

\begin{document}

\preprint{AIP/123-QED}

\title{Laser Absorption Measurements of Electron Density in Nanosecond-Scale Atmospheric Pressure Pulsed Plasmas}

\author{T.Yong}%
\author{A.I.Abdalla}

\author{M.A. Cappelli}
 \altaffiliation{Author to whom correspondence should be addressed:  cap@stanford.edu}
\affiliation{Stanford Plasma Physics Laboratory, Department of Mechanical Engineering, Stanford CA, 94305-3032}
%

\begin{abstract}
We report on time-resolved measurements of electron number density by continuous-wave laser absorption in
a low-energy nanosecond-scale laser-produced spark in atmospheric pressure air. Laser absorption is a result of free-free and bound-free electron excitation, with the absorption coefficient modeled and evaluated using estimates of the time-variation in electron temperature and probe laser absorption path length. Plasma electron number densities are determined to be as high as $n_\text{e}=7\times10^{19}$ cm$^{-3}$, and decay to $1/e$ of their peak values over a period of about 50 ns following plasma formation using a 20 mJ, 10 ns pulse width frequency-doubled Nd:YAG laser. The measured plasma densities at later times are shown to be in reasonable agreement with Stark broadening measurements of the 3s[$^5S{^o}$]-3p[$^5P$] electronic transition in atomic oxygen at 777 nm. This study provides support for the use of such continuous wave laser absorption for time resolved electron density measurements in low energy spark discharges in air, provided that an estimate of the electron temperature and laser path length can be made by accompanying diagnostics.  
\end{abstract}

\maketitle
%

\section{\label{sec:level1} INTRODUCTION\protect}
Current-driven and laser-driven discharges of short duration and relatively low energy ($<$100 mJ) in reactive gases or liquids, i.e., of tens of nanoseconds or less, have attracted considerable attention because they afford the ability to selectively heat electrons leading to desirable chemical kinetics while reducing the degree of background gas heating \cite{chu_2014}. Such low-energy nanosecond-scale plasmas generated in gases at high pressure conditions have been studied for several applications, such as gas reforming \cite{Kogelschatz2003,Bak2015a}, flow actuation \cite{little2012separation}, plasma-assisted combustion \cite{Vincent-randonnier2007,Starikovskaia2015}, and biomedical treatment\cite{Ayan2009}. The appropriate plasma for a particular application is determined by a number of parameters, most important of which are the electron density and temperature ($n_e$ and $T_e$).  The peak electron densities in these plasmas can reach extremely high values as the gas is significantly ionized prior to which significant expansion of the plasma can occur. A convenient means of measuring these parameters at these high pressures is indispensable to the understanding of the kinetics but presents a challenge especially for $n_e$ above 10$^{19}$ cm$^{-3}$.
\par More common methods to measure $n_e$ in plasmas include the use of Langmuir probes \cite{Hopkins2007}, interferometry \cite{ashby1963measurement}, Thomson scattering (TS) \cite{Kempkens2000}, and Stark broadening of spectral line emission \cite{Zhu2009,Orriere2018,Kielkopf2014}. Langmuir probes are relatively straightforward in their implementation but can be perturbing of the plasma for plasma scales comparable to the probe size and the collected current is difficult to interpret in collisional conditions. Thomson scattering, particularly collective TS (CTS) \cite{cameron1996electron}, is a powerful diagnostic tool that can provide good spatial resolution, but its implementation requires an expertise and equipment that is often not available to researchers outside of plasma disciplinary areas. Interferometry, while less demanding than TS in its implementation by comparison, is also a diagnostic that is not often used outside of traditional plasma research laboratories. Optical emission spectroscopy (OES), on the other hand, is widely accessible to non-experts in plasmas. OES is often used to identify species present through their spectral signatures and the broadening of atomic constituent spectral lines due to interactions with free electrons and ions (Stark broadening \cite{griem2012spectral}) constitutes a relatively straightforward way to measure the electron density. However, at very high plasma densities, spectral lines may experience interferences with neighboring lines, blend into the continuum \cite{Bataller2014}, or some are lost altogether due to the lowering of the ionization potential \cite{Hoarty2013}. In nanosecond-scale discharges at high pressures, this precludes the measurements at early times when the plasma densities are quite high but has provided information on the later recombination kinetics of the plasmas \cite{Orriere2018}. Furthermore, self-absorption may become important at high pressures, greatly distorting the spectral profiles \cite{Kielkopf2014}.\\
\begin{figure*}
	\centering
	\includegraphics[width=0.550\textwidth]{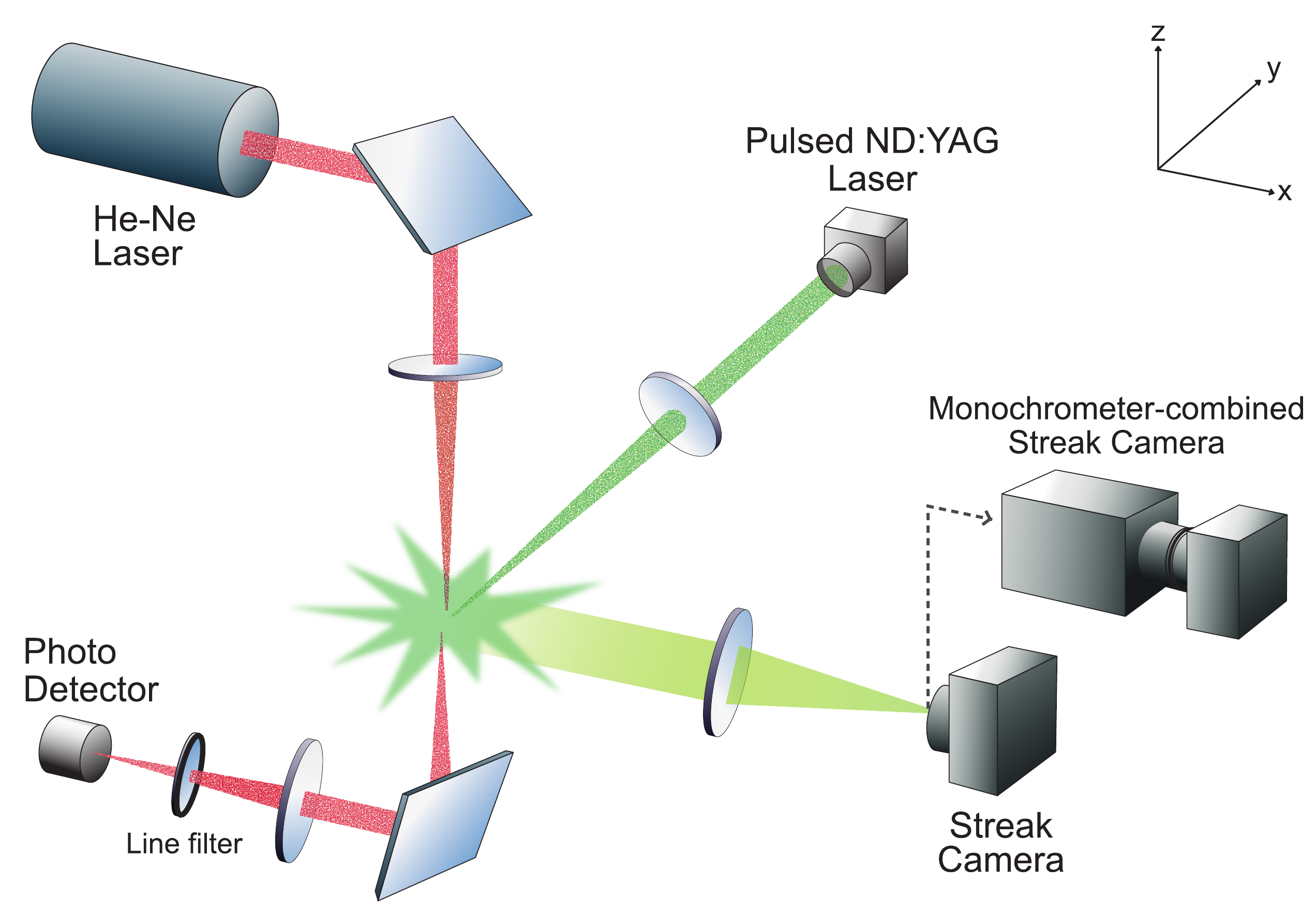}
	\hspace{1cm}
	\includegraphics[width=0.220\textwidth]{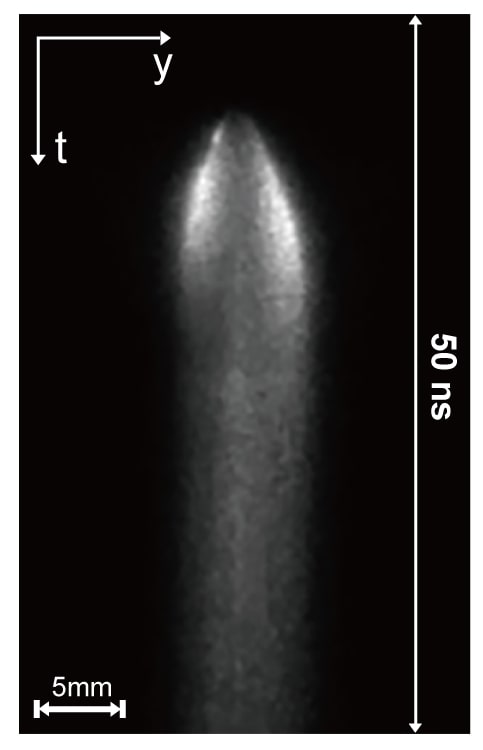}
	\caption{\label{Fig1}(a) Schematic diagram of the experiment. (b) Representative streak image of the plasma emission. The pulsed Nd:YAG laser enters from the right. The slit to the streak camera is parallel to the direction of the incoming pulsed laser.}
\end{figure*}
This paper reports on time-resolved measurements of $n_e$ in a low-energy nanosecond-scale high pressure (1 atm) plasma by continuous-wave (cw) laser absorption. The cw laser is attenuated as a result of free-free and bound-free excitation of the  free and bound electrons, respectively. This method has been used successfully in the past as an electron density diagnostic \cite{Offenberger1971}. Its development in our laboratory is intended for measurements of the time-dependent evolution and decay of $n_e$ in a nanosecond pulse discharge plasma which we, and others, have used for studying chemistry and electron-driven kinetics in reactive gases, \cite{Bak2015a} and nanosecond scale laser discharge plasmas used in plasma-assisted combustion \cite{bak2013studies}. During the initial tens of nanoseconds, high pressure plasmas are sufficiently optically thick to absorb visible light \cite{Bataller2014,Zhang2009}.  Using a fast (sub-nanosecond resolution) and sensitive photodetector, Beer's law allows the extraction of the plasma absorption coefficient, $\kappa$, provided an estimate of the path length through the plasma can be made. As described below, $n_e$ can be determined through a comparison to the absorption coefficient predicted theoretically. The drawback of this measurement, like most absorption-based diagnostics, is that it is a line-of-sight average. Also, an accurate determination of $n_e$ requires an estimate of the electron temperature, $T_e$. The great advantage of the measurement is that it is relatively simple in its implementation and particularly applicable to these high density and relatively small plasmas. At lower plasma densities (e.g., below $10^{17}$ cm$^{-3}$), the absorption is tenuous, requiring a relatively large-scale plasma for good signal-to-noise ratios.

Below, we describe the bread-boarding of this laser absorption measurement on a pulsed laser-produced plasma in air at atmospheric pressure. Such laser-sparks are rich in the spatial-temporal structure, as shown in the study of Harilal et al.\cite{harilal2015lifecycle}, and even more recently, in the computational and experimental studies by a team at the University of Illinois\cite{alberti2019laser,alberti2019modeling,munafo2020computational,nishihara2018influence}.  For laser absorption measurements we use a simple helium-neon (HeNe) laser of low (1 mW) power. For later times in the plasma evolution we compare the measured $n_e$ to that determined from time-resolved Stark broadening of the 3s[$^5S{^o}$]-3p[$^5P$] transition in atomic oxygen (OI) centered at a wavelength of approximately 777 nm. The absorption measurements of $n_e$ make use of previous measurements of $T_e$ in similar laser produced plasma sparks\cite{Borghese1998}. These previous measurements are consistent with an estimate for our plasma based on the absolute intensity and spectral variation in the background continuum over a broad region in the visible range of the spectrum. The electron density measurement also makes use the computed transverse plasma kernel dimensions presented recently by Alberti et al. \cite{alberti2019modeling} As shown below, the plasmas formed by focusing the frequency-doubled output from a Nd:YAG laser of modest power into air reach a peak electron density in excess of $5\times10^{19}$ cm$^{-3}$ and temperatures of $\approx$ 2-4 eV, based on continuum emission averaged over the $\approx$100 ns duration of the luminous plasma. These electron temperatures are also found to be consistent with measured shift-to-width ratios of the OI 777 nm spectral line emission from the central region of the plasma. Imaging of the evolution of the plasma along a direction coincident with the ionizing laser using a streak camera provides insight into its shape, and translating the image of the camera slit along a direction transverse to the laser path provides an experimental estimate of the size of the plasma along the direction of the HeNe probe beam at a time $\approx$20 ns, when the plasma is most luminous. 

\section{Experiment}
The experimental set up is illustrated schematically in Fig 1(a). The 15 Hz pulsed output of a frequency-doubled (532 nm) Nd:YAG laser (Gemini PIV 15, New wave research) with a 10 ns pulse duration (FWHM) and 20 mJ pulse energy is used to produce a laser-breakdown plasma kernel in air. The  5 mm diameter laser beam is focused using a 50 mm focal length plano convex lens to an image distance of 54 mm. The firing of the laser is triggered by the 15 Hz TTL output from a pulse delay generator (SRS model DG535), which also supplies a second TTL pulse to trigger the various plasma diagnostics hardware. The laser absorption is carried out with a continuous wave (cw) 632.8 nm HeNe probe laser (5mW, Melles Griot), 1 mm in diameter, focused through the laser-breakdown plasma at an image distance of 120 mm using a 100 mm focal length plano convex lens to a beam waist of approximately 40 $\mu$m. The precise location of the focus of the probe beam within the plasma is varied to maximize its absorption using a 3-axis micrometer stage on which the focusing lens is mounted. The transmitted probe beam is then re-focused using a second 150 mm focal length lens onto a fast photodiode detector (DET025AL, Thorlabs) which has a 400-1100 nm spectral range sensitivity and a 150 ps temporal resolution. The detector has a built-in lens which focuses the beam onto its active region, 250$
\mu$m in size.  Using a micrometer stage on the focusing lens the transmitted beam is centered onto the active area of the photodiode. The time-varying voltage output of the photodiode is recorded on an oscilloscope, the trace of which is triggered by a second output from the delay generator and averaged (typically over four Nd:YAG laser shots). A narrow pass band filter (centered at 632 nm) is placed between the plasma and the  photodiode detector to minimize interference from the bright, broadband plasma emission. Residual plasma emission still detected by the photodiode is corrected for by recording its small contribution with the HeNe laser blocked. As described below, for a typical laser breakdown plasma the HeNe probe laser is attenuated by as much as 65\%.\\
\begin{figure*}
	\centering
	\includegraphics[trim=0.5cm 6cm 0.5cm 6cm,width=0.48\textwidth]{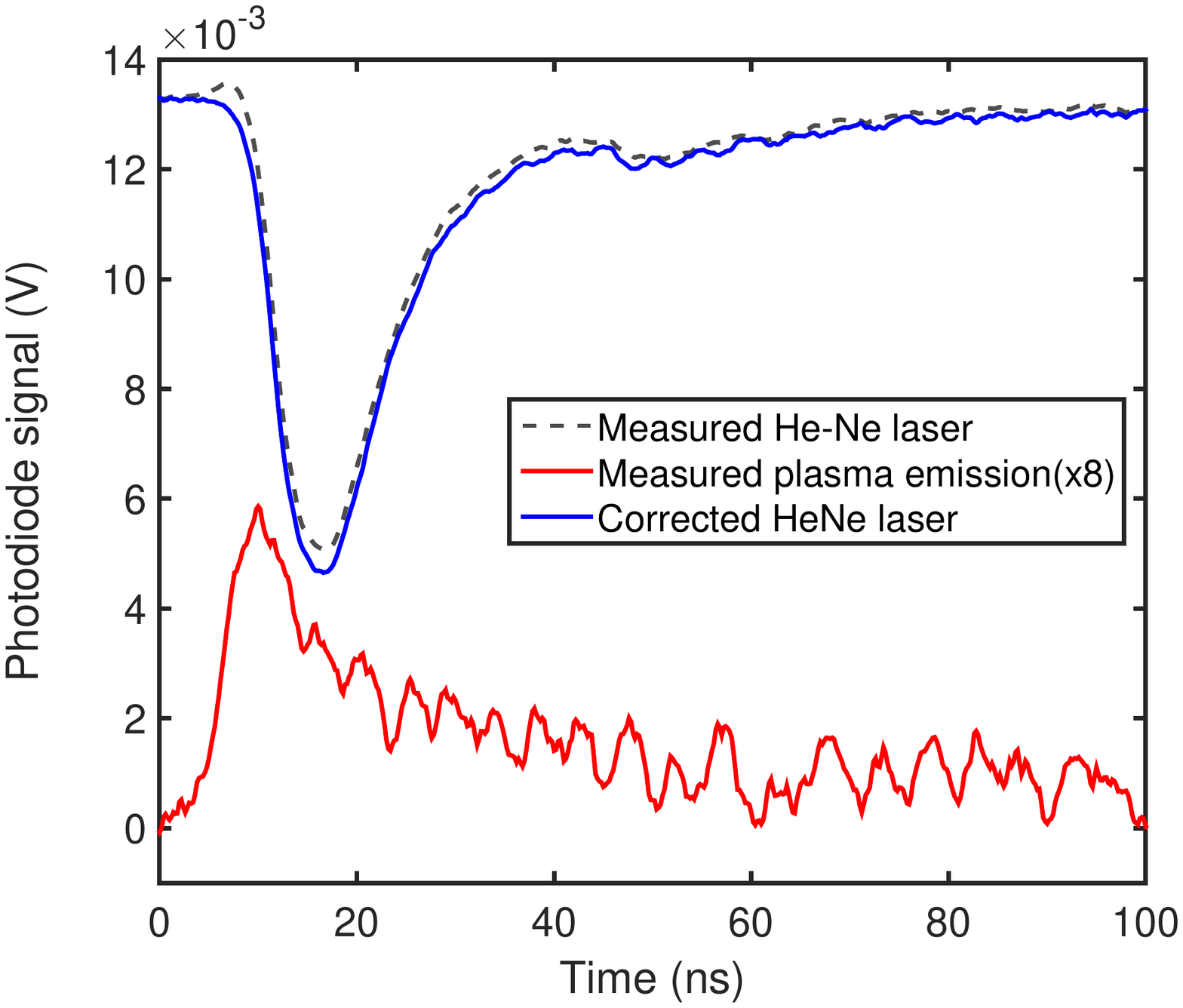}
	\includegraphics[trim=0.5cm 4.5cm 0.5cm 7.5cm,width=0.46\textwidth]{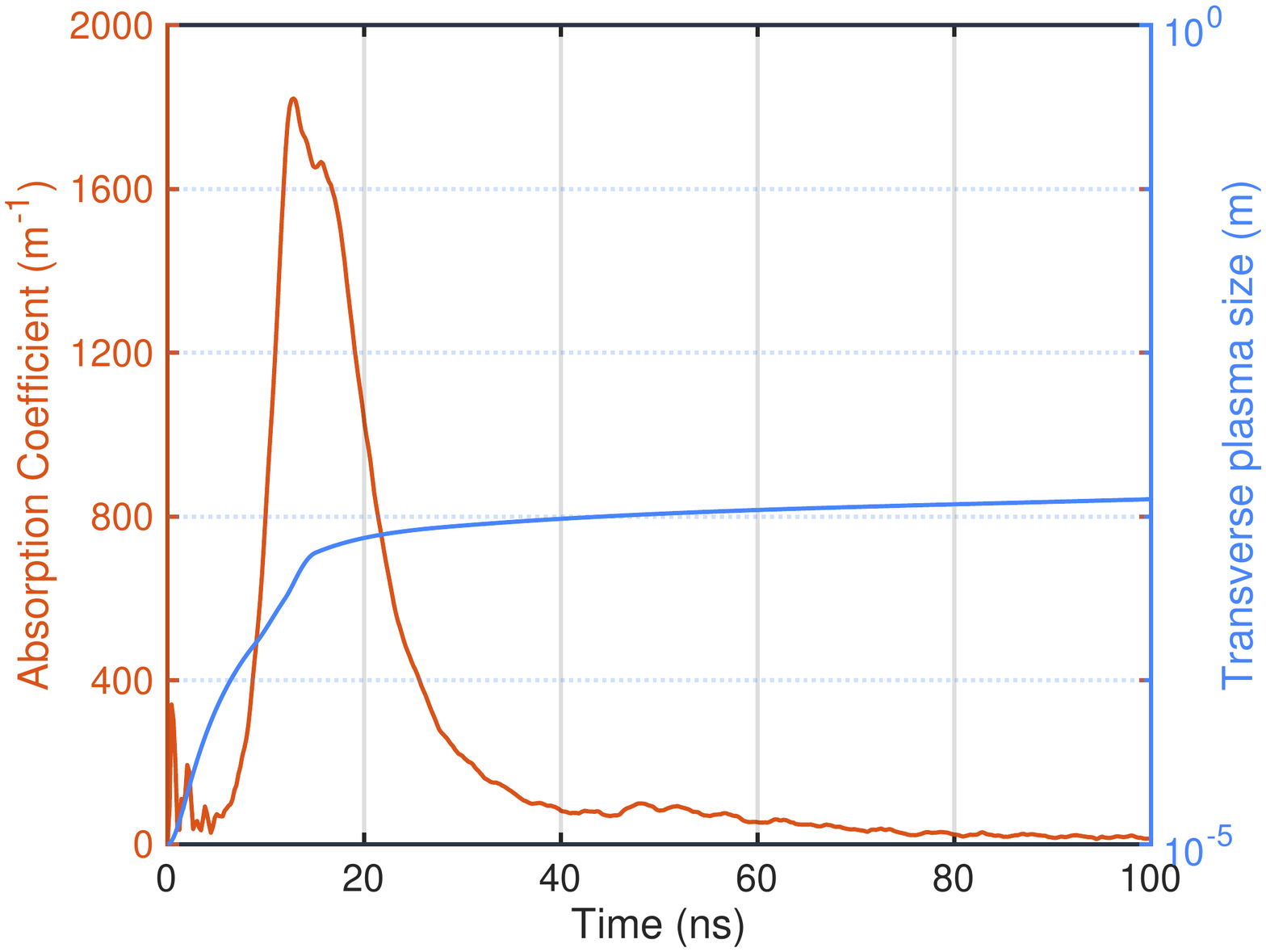}
	\caption{(left)
	Representative photodiode signal recording transmission through the plasma (dashed black curve), corrected for plasma emission (red curve) recorded with the HeNe laser blocked (solid blue curve). (right) Resulting absorption coefficient (red curve) and a fit to the largest transverse dimension of the plasma kernel as computed in the study of Alberti et al\cite{alberti2019modeling} (blue curve).}
\end{figure*}
Time-resolved evolution of the spatial distribution of wavelength-integrated plasma emission along the dimension coincident with the Nd:YAG laser propagation is recorded using a streak camera (Hamamatsu Streakscope Model C4334). The camera is capable of streaking over as small a temporal window of 1 ns with a 2 ps resolution. The streak camera is also triggered by the same delay generator that is used to trigger the photodiode and Nd:YAG laser. The plasma emission is imaged onto the horizontal entrance slit (parallel to the direction of the pulsed laser-forming plasma) using a 100 mm focal length and 26 mm diameter plano-convex lens to a magnification of 0.6 resulting in an estimated spatial resolution of 4.3 $\mu$m. In a typical experiment, the emission is streaked over a period 50 ns with 100 ps resolution. A neutral density filter (0.5 stopping power) is used at the entrance of the streak camera to prevent saturation. A representative streak is shown in Fig. 1(b). Here, the laser arrives from the right hand side. The streak shows an initial rapid expansion of the plasma, which is brightest on both the incoming and trailing side of the laser, qualitatively similar to the results of Nishihara et al \cite{nishihara2018influence}.  In some experiments, the streak camera is coupled to the output of an optical monochromator (Hamamatsu Model C5094 1/4-m focal length, f/d = 4). The monochromater collects the emission using fibre-coupled optics to image the plasma onto its 50 $\mu$m vertical entrance slit. A fibre lens assembly serves to image the plasma emission onto the horizontal entrance slit of the streak camera which is  mounted to the exit plane of the monochromator. The lens assembly is adjusted to optimize the signal, imaging the brightest (and hence hottest) region of the plasma. The streak camera sweeps the spectral emission over a spectral window of 20 nm. This allows the recording of the Stark-broadened OI line at 777 nm. For results described here, the spectra are streaked over a period of 100 ns time duration with 200 ps time resolution. In both cases, the collection optics is aligned such that the image of the streak camera entrance slit is coincident with the centerline of the incident Nd:YAG laser.
Finally, for an estimate of the electron temperature, we image the the brightest region of the plasma using optical fiber coupling onto the entrance of a compact optical spectrometer (Ocean Optics), calibrated for absolute and relative spectral response. However, the compact monochromator averages the signal over a 1 ms time window.  The resulting spectra are therefore more representative of an average over the period of strongest (and presumably highest temperature) emission typically, 50-100 ns in duration. To estimate the absolute intensity of the continuum, we correct the spectrometer signal for the ratio of the plasma duration to spectrometer integration time.    
\section{Results}
A representative temporal photodiode signal from the cw HeNe probe laser is shown as a dashed black line in Fig. 2(a). In a separate shot, with the probe laser blocked, we record the broadband plasma emission on the same photodiode (solid red line, expanded by 8$\times$) to correct the probe laser trace for background interference, which causes the inflection seen near the onset of the drop in the transmitted probe laser signal. It is noteworthy that the peak in the plasma emission occurs approximately 8 ns before the maximum attenuation in the probe laser signal suggesting that for early times, the plasma is brightest before reaching its most strongly ionized state. The emission-corrected probe laser signal, $\mathit{I}$, is the solid blue line in Fig. 2(a). In the corrected signal, we see that the onset of the attenuation is monotonic, as expected. It is apparent that the plasma becomes quite opaque with a maximum attenuation in the probe laser of about 65\%. The temporal variation in the transmission in probe laser intensity, $\mathit{T=I/I_{o}}$, allows us to determine the plasma absorption coefficient, $\kappa$, through Beer's law,
\begin{align*}
 \kappa =-\frac{1}{L} ln\left (\frac{I}{I_{o}}  \right )\label{eq5)}\tag{1}.
\end{align*}
Here, $\mathit{I_{o}}$ is the probe laser intensity (proportional to the recorded photodiode signal) in the absence of the absorbing plasma and $\mathit{L}$ is the path length traversed by the probe through the plasma, which varies in time as the plasma forms and expands.  A representative plasma absorption coefficient (at 632.8 nm) is
shown as the red line in Fig. 2(b). Qualitatively, we see that the plasma is most opaque between approximately 10 and 20 ns, representing the period in time when the plasma is most dense as the attenuation is a consequence of electron free-free and bound-free absorption. An analyses of absorption data based on experimental and computational measurements of electron temperature and probe laser path length provides determination of the electron number density. In our studies, the path length is estimated at a time when the plasma emission is most luminous, i.e., at $ \approx $ 20ns, by varying the image of the horizontal entrance slit of the streak camera onto the plasma until the emission was no longer visible in the streak. At this time, we estimated that the size of the plasma along the direction perpendicular to the incoming ionizing laser is $L\approx$ 0.9  mm. As shown in recent computational studies\cite{alberti2019modeling}, the shape of the plasma varies considerably. In the first $\approx$30 ns the plasma kernel increases quickly in size as a result of laser heating and expansion, and less so at later times due to the increasing importance of plasma recombination. We also include in Fig. 2(b) the transverse plasma size, $L(t)$, at its widest location, using a fit to the computational data of Alberti et al\cite{alberti2019modeling}.

In general, absorption by the free electrons in the plasma includes contributions to the absorption coefficient from bound-bound transitions ($\kappa_{bb}$), bound-free (electron-ion) interactions ($\kappa_{bf}^{ei}$), as well as free-free electron-ion ($\kappa_{ff}^{ei}$) and electron-neutral ($\kappa_{ff}^{en}$) interactions. The total spectral absorption coefficient for these interactions can be expressed as:\\
\begin{align*}
\kappa=\kappa_{bb}+\kappa_{bf}^{ei}+\kappa_{ff}^{ei}+\kappa_{ff}^{en} \label{eq2}\tag{2}.
\end{align*}
The spectral absorption coefficients can be obtained from available theoretical expressions for the corresponding spectral emission coefficients, $\epsilon(\lambda)$, using Kirchoff's Law,\\
\begin{align*}
\frac{\epsilon(\lambda)}{\kappa(\lambda)}=B_\lambda(\lambda,T)=\frac{2hc^2}{\lambda^5}\frac{1}{\exp{\left(\frac{hc}{\lambda kT_e}\right)}-1}\label{eq3}\tag{3}
\end{align*}
where $B_\lambda$ is the spectral energy radiance of a black-body. Expressed in this way,  $h$, $c$, $\lambda$, and $k$, represent Planck's constant, the speed of light in vacuum, wavelength, and the Boltzmann constant, respectively. We assume that free and bound states are redistributed by collisions with free electrons at a temperature, $T_{e}$, as over this short period of time, we expect the electrons to be hotter than the heavy particle temperature as little electron-ion energy transfer will take place. Since we expect the ionization fraction to be high and that free electron-ion interactions are much stronger than free electron-neutral atom interactions \cite{hughes1975plasmas}, we take as a consequence,
\begin{align*}
\kappa_{ff}^{ei}>>\kappa_{ff}^{en}
\end{align*}

This assumption will be revisited and examined in the discussion in Sec. IV. The laser probe wavelength (632.8 nm) is relatively free of nearby bound-bound transitions in laser plasma sparks in air\cite{harilal2015lifecycle} so we take $\kappa_{bb} \approx$ 0. The corresponding remaining contributions to the spectral emission coefficient are obtained from Venugopalan \cite{venugopalan_1971},\\
\begin{align*}
\epsilon_{ff}^{ei}(\lambda)&=\sum_s{\frac{C_1n_en_s}{\lambda^2T_e^{\frac{1}{2}}}z_s^2\left[G_s\exp{\left(\frac{-hc}{\lambda kT_e}\right)}\right]}\;\;\;\;\\
\epsilon_{fb}^{ei}(\lambda)&=\sum_s\frac{C_1n_en_s}{\lambda^2T_e^{\frac{1}{2}}}z_s^2\cdot\frac{g_{s,1}}{U_{s}(T_e)}\left[\xi_s\left(1-\exp{\left(-\frac{hc}{\lambda kT_e}\right)}\right)\right]\label{eq4}\tag{4}
\end{align*}
where C$_1=1.630\times$10$^{-43}$ Wm$^4$K$^{\frac{1}{2}}$sr$^{-1}$. Here, z$_s$, $g_{s,1}$ and $U_{s}(T_{e})$ represent the charge,  ground-state degeneracy and partition function of the of $s-$ionic species, and the summation extends over the ionized species in the air mixture. The parameters, $G_{s}$ and $\xi_{s}$ are the Gaunt and Bieberman factors that account for non-hydrogenic effects which are species dependent. The Gaunt factors, $G_{s}$, are taken to be approximately unity regardless of species, which is seen to be a good approximation for a broad range of electron energy ($kT_e$) and incident photon wavelength expected in our studies \cite{Sutherland}. On the other hand, $\xi_s$ can vary substantially with the ion species and while there are estimates of its value for atomic ions, there are very few studies on its determination for molecular ions. In the studies of Bibermann et al., $\xi_s$ ranges from about 0.5 to 1.5 for a number of atomic ions \cite{Biberman_1967}. We use this range of values for estimating the bound-free contribution to the total absorption, assuming that the most abundant ionized species is singly-ionized atomic nitrogen ($N^{+}$), consistent with the studies of Orriere et al \cite{Orriere2018}. This assumption reduces the summation in Eq.\eqref{eq4} to a single contribution from $N^{+}$, and the absorption coefficient becomes:\\  
\begin{align*}
&\kappa(\lambda)=\frac{C_1n_e^2\lambda^3}{2hc^2T_e^{\frac{1}{2}}}\times\\
&\left[1-\exp{\left(-\frac{hc}{\lambda kT_e}\right)}+2\xi_{N^{+}}\frac{g_{N^{+},1}}{U_{N^{+}}}\left(\cosh{\left(-\frac{hc}{\lambda kT_e}\right)}-1\right)\right]\label{eq5}\tag{5}
\end{align*}
Studies of similar pulsed laser-produced air sparks \cite{Hohreiter2004,ELSHERBINI2006,Yalcin1999} have reported $n_e \approx$ 2$\times$10$^{18} $cm$^{-3}$ and $T_e$ $\approx$ 2 eV at a time of 100 ns after plasma initiation. This value of temperature helps to establish the range expected in our studies, and a means of estimating the contribution of the two terms in parenthesis in Eq.\eqref{eq5} above, i.e., the free-free absorption (first term) and bound-free absorption (second term) in the equation. With these values, we can estimate the absorption coefficient for the 632.8 nm wavelength of the probe laser beam. The free-free contribution factor (the first term in parenthesis, $f_{ff}$) and its dependence on $T_e$ is shown in blue in Fig. 3, whereas the multi-colored lines are that for the bound-free factor (second term in parenthesis, $f_{bf}$), for a range of Biberman factors. For these calculations, we use the $N^{+}$ partition function tabulated by Capitelli et al.\cite{capitelli2005tables} We see that for temperatures below $\approx$ 0.3 eV, the bound-free contributions dominate and there is a variation in values within the range of Biberman factors considered. Above 3 eV, the contributions are dominated by the free-free transitions. For the range of temperatures of 2 - 4 eV, we expect the absorption to be dependent on the treatment of the bound-free contribution at the lower end, and entirely from free-free electron transitions facilitated by $N^{+}$ at the higher end of the temperature range.

\begin{figure}
	\centering
	\includegraphics[trim=0.5cm 6cm 0.5cm 6cm,width=0.48\textwidth]{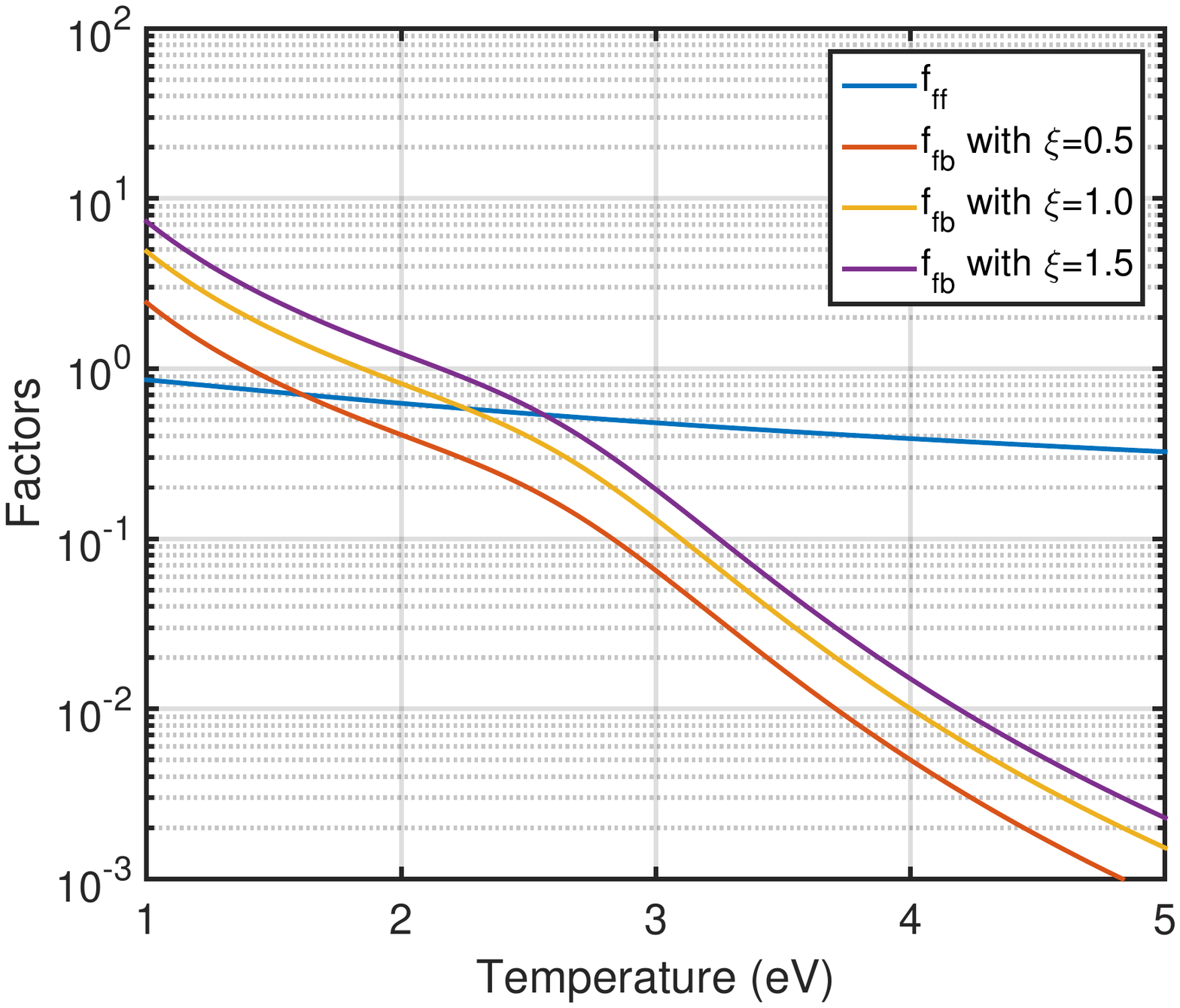}
	\caption{Contributing factors of free-bound and free-free transitions to the total absorption coefficient for N$^+$ taken to be the dominant ion in the plasma. Bound-free factors are evaluated for a range of Biberman factors.}
\end{figure}

\begin{figure}
	\centering
	\includegraphics[trim=0.5cm 4.5cm 0.5cm 4.5cm,width=0.46\textwidth]{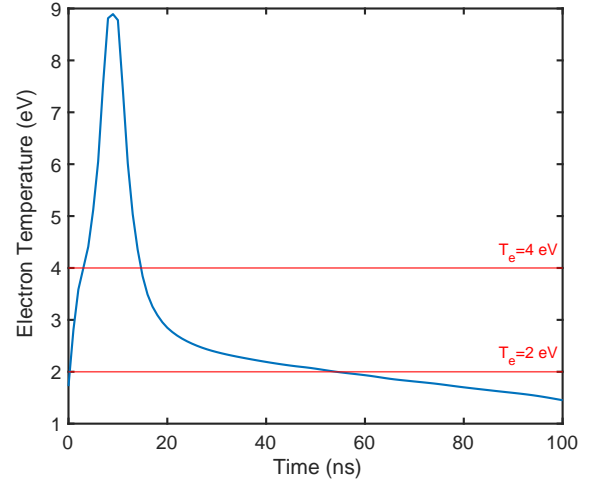}
	\caption{Temporal variation in electron temperature of a laser-produced spark. The blue line represents a fit to the experimental data of Borghese\cite{Borghese1998}. Also shown as the horizontal red lines represent the lower and upper values for our electron temperature estimated from continuum measurements averaged over the duration of strong plasma emission.}
\end{figure}

Since the absorption coefficient is sensitive to $T_{e}$, an estimate of the electron temperature is beneficial, including its temporal variation. For similar laser-produced plasmas in air, the temporal variations in $T_{e}$ have been reported \cite{Borghese1998}. It tends to peak at values of  $T_{e}\approx$ 9 eV at 10 ns following plasma initiation, and drops first quickly, and then gradually to $T_{e}\approx$ 1 - 2 eV over a range of about 100 ns. As we have not measured the temporal variations in $T_{e}$ for our plasmas we use the measurements reported by \cite{Borghese1998} to reduce our measured time-resolved absorption coefficients to electron number density. A reproduction of this temperature variation with time is shown in Fig. 4. As a check on this data, we have carried out measurements of time-averaged relative and absolute continuum emission for comparison.  Although spectra recorded with the compact spectrometer are time-averaged over the duration of the emission, as mentioned earlier, the temporal variation in the emission recorded by the fast photodiode suggests that this emission lasts for approximately 100 ns. This value, together with an absolute intensity calibration of the spectrometer affords a confirmation on the temperature (averaged over 100 ns), for self consistency. We plot the measured spectral radiance of the plasma over a broad wavelength range in Fig. 5. Also plotted is the computed total continuum blackbody emission for electron temperatures ranging from 2-4 eV. Our measured time-averaged temperatures are biased somewhat lower than the temperatures expected at early times, somewhere near the peak in plasma emission (10-20 ns). We see that a temperature of $T_{e} \approx$ 2 - 4 eV describes both the absolute intensity as well as the measured wavelength variation in the underlying continuum reasonably well. This range of temperature is compared to that of Borghese et al. \cite{Borghese1998} in Fig. 4. It is expected that our measured average temperature is somewhere above the lowest and below the peak in the time-resolved data. 

\begin{figure}
	\centering
	\includegraphics[trim=0.5cm 6cm 0.5cm 6cm,width=0.48\textwidth]{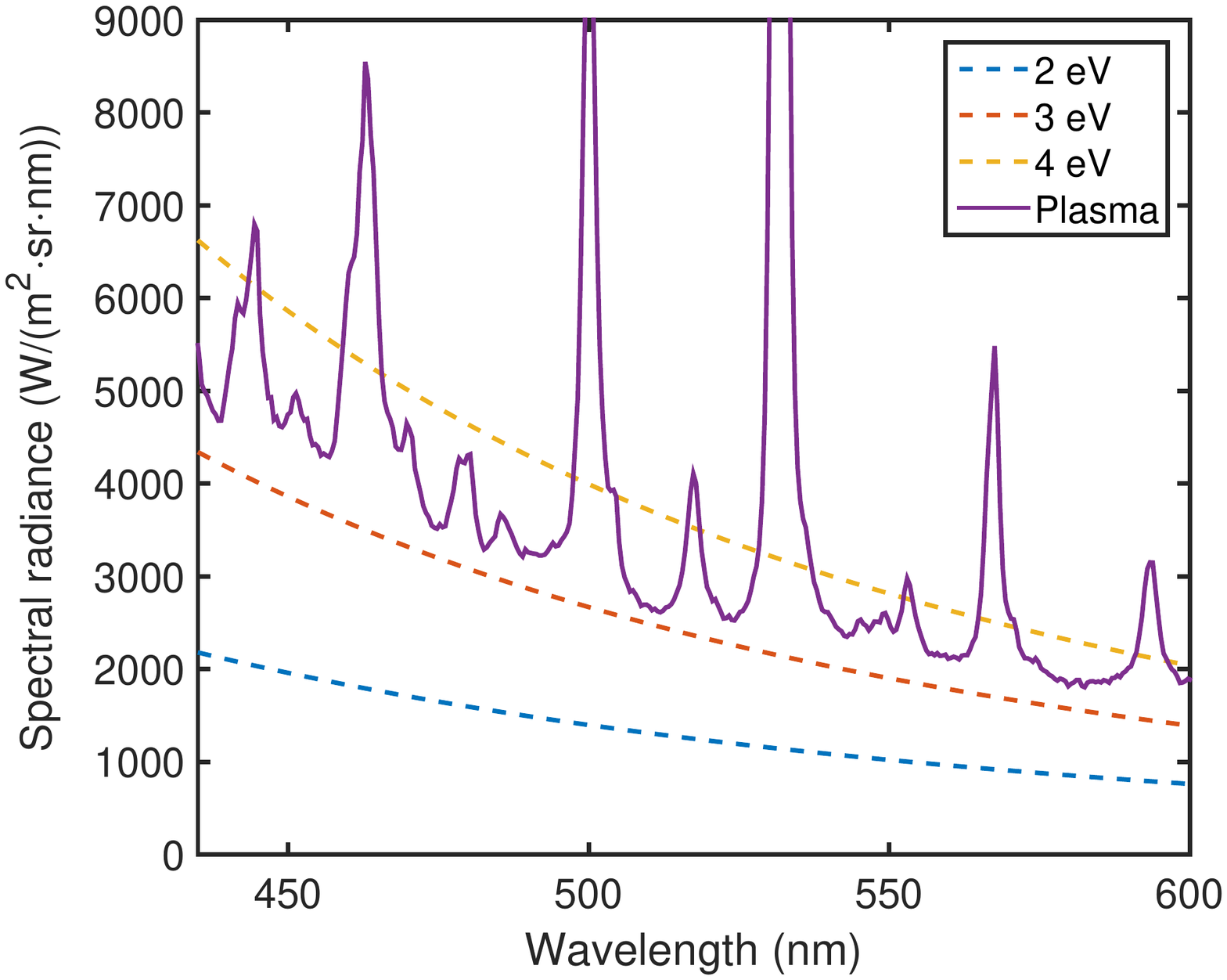}
	\caption{Comparison in the measured and computed absolute continuum intensity, assuming a black-body radiation spectrum.}
\end{figure}

Using the $T_{e}(t)$ shown in Fig 4, and, for the free-bound contribution to absorption, a Biberman factor of 1, we now evaluate the absorption coefficient and hence the temporal variation in the electron number density, $n_{e}$. The result is displayed as the blue line in Fig. 6. The error bars represent our estimated range of uncertainty in these values. 

We see that $n_{e}$ peaks at a density of approximately 7 $\times$ $10^{19}$  cm$^{-3}$. The peak exceeds, by more than a factor of two, the Loschmidt number. We attribute some of this to the dissociation of the diatomic species as the temperature rises substantially in the first approximately 10 ns. A further elevation above the Loschmidt number can also be a result of multiple ionization. We also see that at 100 ns, the plasma density falls to about 2 $\times$ $10^{18}$ cm$^{-3}$ approaching our estimated noise floor of about $10^{18}$ cm$^{-3}$, limited mainly by Nyquist (Johnson) noise associated with thermally-induced fluctuations in the photodiode current. Finally, we note in the electron density the presence of temporal fluctuations that are quite pronounced at later times.  The frequency of these fluctuations is approximately 0.3 GHz. These fluctuations are also seen in the detected plasma emission trace, as evident in Fig. 2.  There have been reports of propagating striations of 0.1 mm scale in similar laser-induced sparks through schlieren imaging\cite{harilal2015lifecycle}. As reported in that study, the origin of such features is still not clear, however, it is noteworthy that the spatial and temporal scales associated with these fluctuations suggest disturbances that propagate close to the ion sound speed. 

As a second measurement of electron density we perform time-resolved emission measurements of the Stark broadening of the 777 nm atomic oxygen line, which is somewhat intense relative to the background continuum radiation so that $n_e$, particularly at early times, can be determined with reasonable accuracy. The recorded spectra are fit to Voigt profiles to distinguish the Lorentzian width of the Stark broadened lines from the Gaussian contribution ($w_G)$ from both instrument ($w_I$) and Doppler broadening ($w_D$), given by:

\begin{figure}[!b]
	\centering
	\includegraphics[trim=2.5cm 0.5cm 2.5cm 0.5cm, scale=0.375]{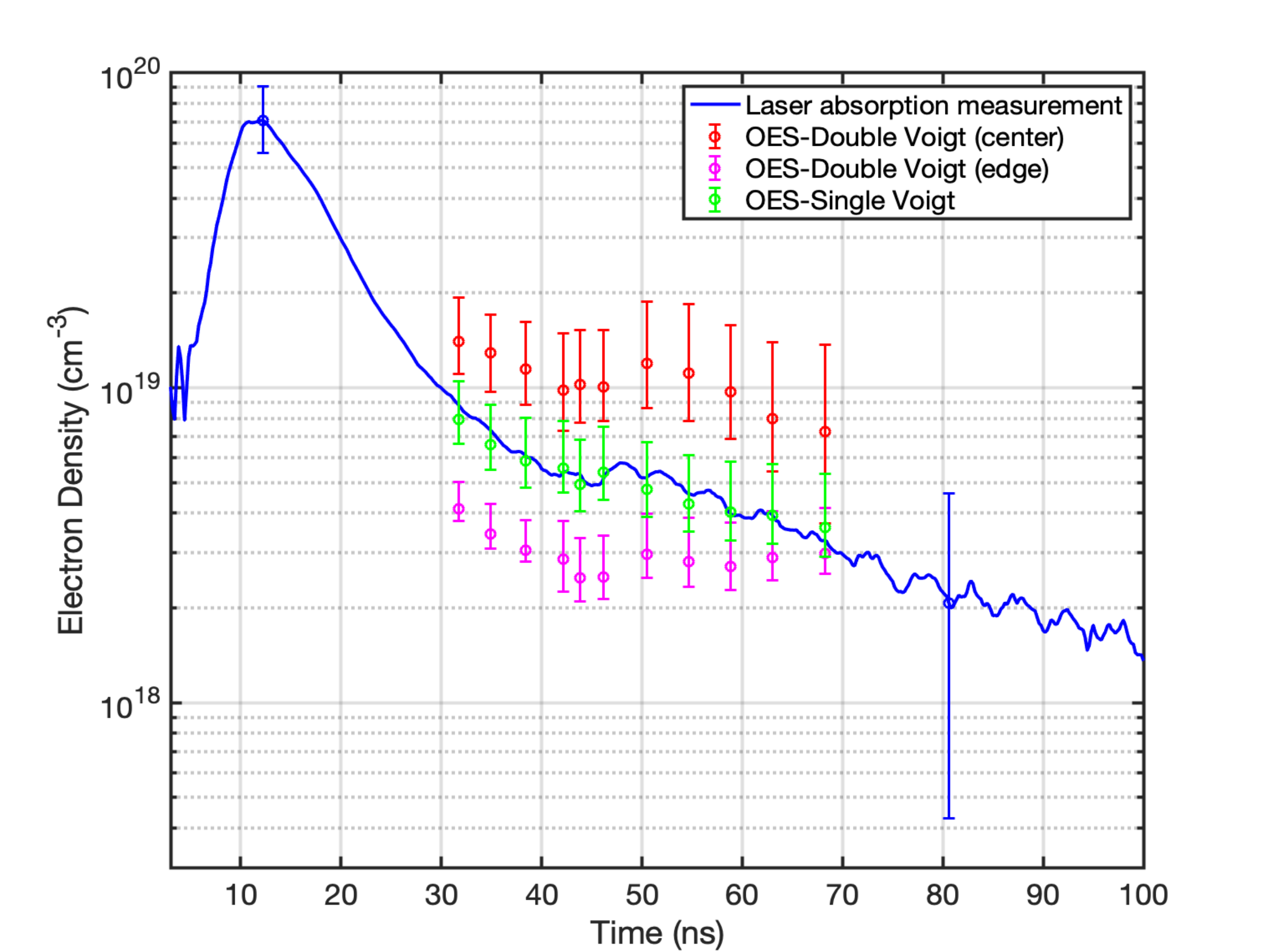}
	\caption{A comparison of the temporal variation in $n_{e}$ measured by absorption (blue line) to that measured using optical emission and the Stark broadening of the oxygen 777 nm electronic transition.}
\end{figure}

\begin{align*}
w_{G}&=\sqrt{w_{D}^2+w_{I}^2}\\
w_{D}&=7.17\times10^{-7}\times\lambda_o\sqrt{\frac{T}{M}}
\end{align*}

Here $\lambda_o$ is the center wavelength of the line, $T$ is the translational temperature of atomic oxygen in K, and M is its atomic mass in atomic mass units. As an estimate to evaluate the importance of Doppler broadening, we use $T$=10,000 K, resulting in $w_D\simeq$ 0.014 nm. Although significant, this broadening is much less than that measured for our instrument resolution, even at translational temperatures approaching that of the electrons. A spectral scan of the emission from a mercury lamp indicates that $w_I$ is approximately 0.36 nm. The resultant $w_G$ combined with the measured spectral width ($w_{\text{meas}}$) is used to deconvolve and determine $w_L$, which is mainly due to the electron-impact collisional Stark broadening width,$w_S$, corrected for quasi-static ion microfields\cite{griemplasma}:
\begin{align*}
w_{S}(n_e,T_e)=2w_e(T_e)\times\left[1+1.75\alpha_e(T_e)(1-0.75R_e)\right]\frac{n_e}{n_{eo}}
\end{align*}
with
\begin{align*}
R_e=8.99\times10^{-2}\frac{n_e^{1/6}}{T_e^{1/2}}.
\end{align*}
Here, $w_e$ is the electron Stark half-width (in Angstroms), $\alpha_e$ is a dimensionless parameter that accounts for the effect of ion microfields, $R_e$, also dimensionless accounts for the Debye shielding of the ion perturbers by the free electrons, and  $n_{eo}=10^{16}$ cm$^{-3}$ is the reference electron density for the tabulated values of $w_e$. The estimated range of $T_e$ from the continuum radiation spectrum of Fig. 4 is used to evaluate $w_e$ and $\alpha_e$. At these high densities, temperatures, and corresponding pressure, we should consider the impact of van der Waals broadening on the OI 777 nm transition lineshape. Using the broadening theory described by Griem\cite{griemplasma}, and the specific reduction for the 777 nm line as presented by Laux\cite{laux1993optical}, the van der Waals broadening rate (in Angstroms) is approximately $\Delta \lambda _{vdW}=\left ( 11.2-2.12\times10^{-4}T \right )T^{-0.7}p$. Here, $T$ is the gas temperature (in Kelvin) and $p$ is the gas pressure (in atm). We take, as a conservative estimate, $T$=3 eV and a number density of $7\times10^{19}$ cm$^{-3}$, and find $\Delta \lambda _{vdW}$=0.08 nm, confirming that van der Walls broadening of this OI transition is relatively small in comparison to other broadening mechanisms for our plasma conditions.

\begin{figure*}
	\centering
	\includegraphics[width=0.47\textwidth]{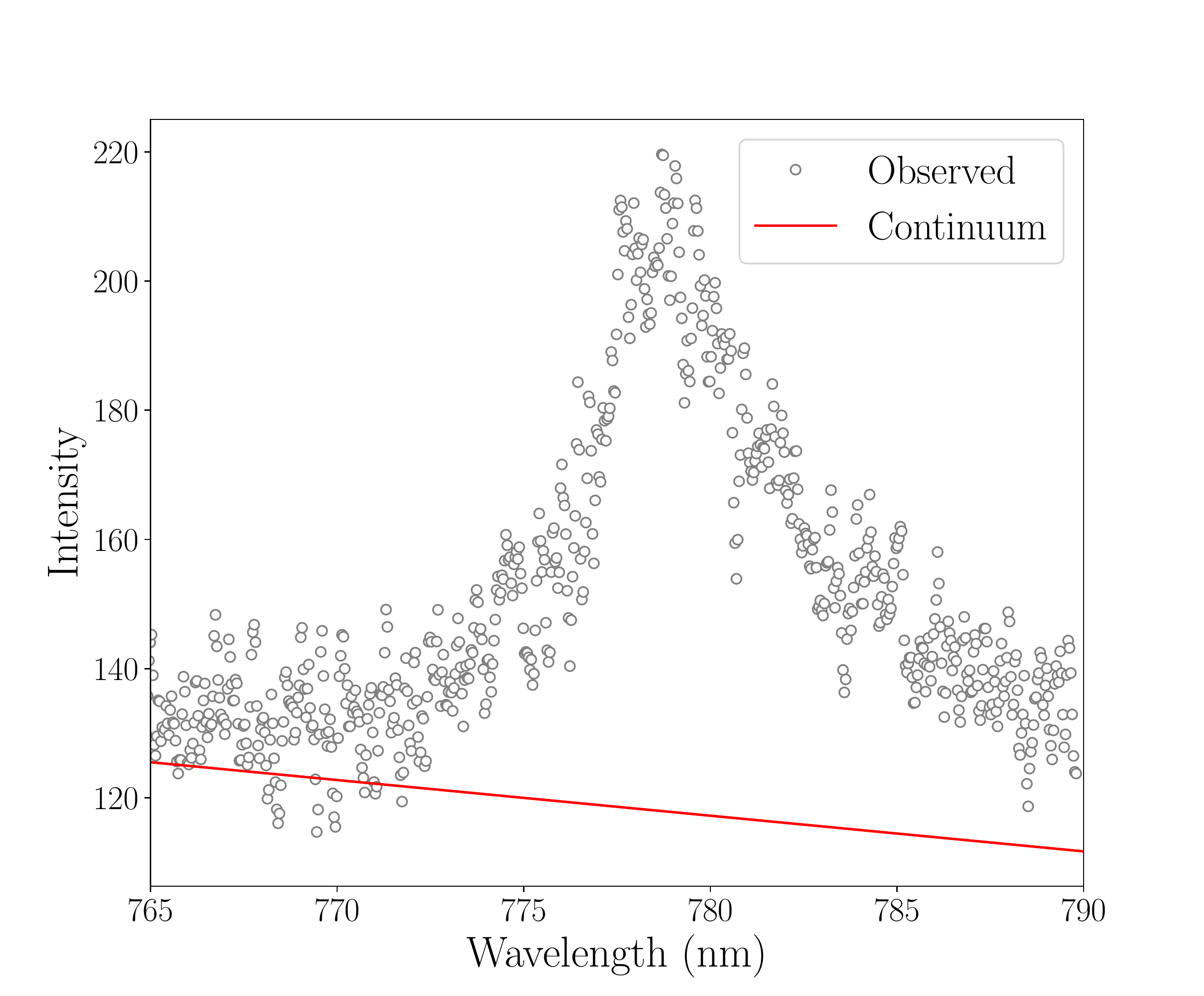}
	\includegraphics[width=0.47\textwidth]{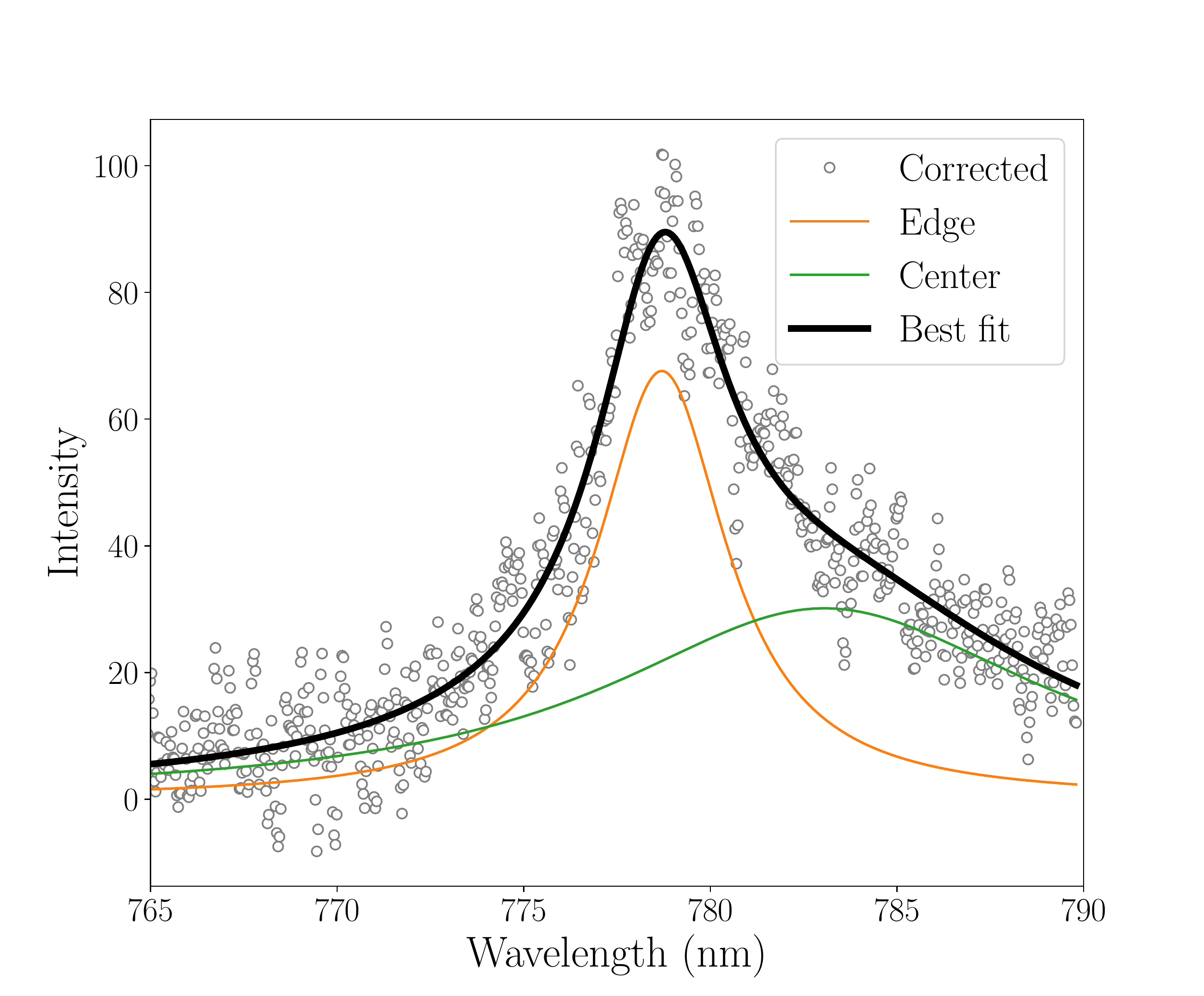}
	\caption{(a) The observed emission spectrum (white-filled circles) and scaled 3eV continuum background (red line). (b) Corrected spectrum (white-filled circles) and fitted Voigt profiles for the center (green) and edge (orange) of the plasma, and the resultant fit of summing the two Voigt profiles.}
\end{figure*}
Fig. 7(a) shows a representative 777 nm emission spectrum (white-filled circles) observed early in the evolution of the plasma kernel (30 ns). Also shown is an underlying red curve, the slope of which represents the continuum radiation calculated based on values of $T_e$ = 3 eV, scaled to blend with the lowest wavelength limit of the spectral scan. This corrected spectrum using the  $T_e$ = 3 eV background continuum is shown In Fig. 7(b) (also white-filled circles). The black line shown is a double-Voigt profile fit to account for the possibility of a higher density (and hotter) plasma core (center), surrounded by a lower density (and cooler) plasma corona (edge), as it is apparent that a single-Voigt profile would not be suitable due to the distorted high wavelength side attributed to a non-uniform plasma density distribution. The best fit, obtained by a regression analysis, consists of a wider and more shifted Voigt profile (green curve) together with a more narrow and less shifted profile (red curve). The values of $n_e$ obtained from the Lorentzian contributions to the two-Voigt fit gives a center and edge value of $n_\text{e,center}=1.18\times10^{19}$ cm$^{-3}$ and $n_\text{e,edge}=2.64\times10^{18}$ cm$^{-3}$, respectively.

For comparison, Fig. 6 also includes in addition to the measured electron number density from the laser absorption, the electron number density inferred from the Stark-broadened OI line. The red and magenta circles represent the plasma values inferred from the double-Voigt fits, with $n_{e,center}$ and $n_{e,edge}$, respectively. We also show the results of a single-Voigt fit (green circles), which tends to lie between the other two, as expected. The data shows a good agreement with those obtained from the HeNe laser absorption, and seem to agree best with the data extracted from the single Voigt fit analysis although all three analyses are likely to be within the range of the uncertainty in the laser absorption measurements at these later times. For earlier times, i.e., before 30 ns, the Stark broadening is extremely wide and the spectral line blends into the continuum, precluding its use as a definitive measurement of $n_e$.  It is noteworthy that for nearly the entire range of measurements extracted from the Stark-broadened OI line, the Stark shift-to-width ratios for this transition\cite{griemplasma} estimated from the contribution to the profiles from the plasma core result in values of $T_{e}$ $\approx$ 3 eV, consistent with the continuum measurements described above and those of Borghese et al\cite{Borghese1998}.  This estimate, as well as that of the electron density, is somewhat sensitive to the determination of the continuum background placement in Fig. 7(a), although its slope is constrained by the temperature of $T_{e}$ = 3eV. 

\section{discussion and summary}
The results described here indicate that laser absorption, primarily through free-free (inverse Bremsstrahlung) and bound-free electron-ion interactions affords a convenient and accurate way to study the early electron number density dynamics of a low-energy laser spark plasma. We believe that this measurement can also be applied to the measurement of electron density in plasmas formed by pulsed nanosecond-scale electric discharges. In its implementation it requires a high speed photodiode detector and a low power cw probe laser, preferably one that has high wave front quality and low divergence, such as a HeNe gas discharge laser. The temporal variation of $n_e$ is extracted directly from the laser attenuation given a determination of the absorption coefficient using an estimate of the electron temperature and probe laser absorption path length. This measurement has advantages in relatively high density laser or discharge spark plasmas at early times during the plasma formation where the analysis of the emission spectra is made difficult as a result of overlapping and/or blending of emission lines into the continuum. Furthermore, while both methods may be line-of-sight based, the use of Stark broadened emission lines from neutral species for extracting n$_e$ is further complicated by the fact that highly-ionized regions of the plasma, i.e., central regions of high plasma density, are less emitting as this region of high temperature experiences neutral species burnout.

In our theoretical determination of the absorption coefficient, $\kappa$, we have assumed that the contribution of free-free electron-ion interactions dominate over electron-neutral interactions. To determine this contribution to the absorption we use the volume emission coefficient given by Venugopalan \cite{venugopalan_1971},
\begin{align*}
\epsilon_{ff}^{en}(\lambda)&=\frac{C_2n_en_aT_e^{3/2}}{\lambda^2}\times\\
&\left[\left(1+\frac{hc}{\lambda kT_e}\right)^2+1\right]Q^{en}(T_e)\exp{\left(-\frac{hc}{\lambda kT_e}\right)}\label{eq5}\tag{6}
\end{align*}
with C$_2$=1.026$\times10^{-34}$ Wm$^2$K$^{-\frac32}$sr$^{-1}$. Here, $Q^{en}$ is the average electron-neutral momentum scattering cross-section, which is dependent on $T_e$. At early times (10-30 ns), we assume based on the high electron number densities estimated, that the plasma is fully ionized, decreasing in number density to $\approx 10^{19}$cm$^{-3}$ largely because of the initial expansion \cite{alberti2019modeling}.
For later times, the density falls from $10^{19}$cm$^{-3}$ to about 2$\times$$10^{18}$cm$^{-3}$ through further expansion and electron-ion recombination. As a worst case, assuming that the drop in electron density beyond 30 ns is solely due to recombination, then at these later times we might have a more weakly ionized plasma with $n_{n}/n_{e} \approx 5$. To estimate the potential impact that  $\kappa_{ff}^{en}$ may have on the inferred electron number density we include its contribution using Eqn. 6, assuming a total electron-neutral momentum scattering cross section,\cite{neynaber1963low, thomas1975low} $Q^{en}\approx 10^{-20}$ m$^{2}$. In doing so, we consider values of $n_{n}/n_{e}$ = 0, 1, and 10. $n_{n}/n_{e}$ = 0 represents neglecting the electron-neutral free-free interactions, whereas  $n_{n}/n_{e}$ = 1 and $n_{n}/n_{e}$ = 10 represent a moderate degree of ionization, and a very low degree of ionization respectively. We find that with Biberman factors near unity, the electron-neutral contributions have little effect on the measurements at later times (because of the relatively low electron temperature). At early times, the peak electron number density falls to $n_{e}$ = 6$\times$10$^{19}$ cm$^{-3}$ and 5$\times$10$^{19}$ cm$^{-3}$ with $n_{n}/n_{e}$ = 1 and 10, respectively. As a result, we believe that for our conditions, neglecting the electron-neutral contributions to the free-free absorption coefficients is justified.  However, more accurate values that these contributions make to the overall free-free absorption would require accurate data for the momentum scattering cross sections for these relatively low temperatures, and a determination of the neutral atom density, which is strongly affected by the complex gas dynamics. 

Our measurements of $n_{e}$ are also predicated on measurements of $T_{e}$, which we have made only indirectly through studies of the continuum emission averaged over a period of time in which the emission is strongest. Our measured Stark broadening shift-to-width ratios do confirm the relatively low values of $T_{e}$ in the range of t = 30-80 ns, consistent with past measurements of laser plasmas in air\cite{Borghese1998}. To analyze our experimental measurements at times during the plasma formation and subsequent early expansion phase, we used the results of Borghese et al. \cite{Borghese1998}, which, although similar in its experiment to ours, was at slightly higher laser pulse energy and at the Nd:YAG fundamental wavelength (1064 nm). A strongly-varying temperature during these early times seems to have a significant impact on the inferred temporal variation of $n_{e}$. A temporally-resolved measurement of $T_{e}$ for our experimental conditions is desirable, and future experiments will explore the possibility of using a second, lower energy and shorter duration (picosecond) pulsed laser to evaluate and confirm both the time-dependent $T_{e}$ and $n_{e}$ by Thomson scattering.

Finally, we note that while the measurement of the absorption coefficient does provide a quantitative measure of the average of the the electron number density-path length product, a measure of the electron density provides a more meaningful determination of the plasma state. An accompanying and accurate measurement of $L(t)$ is necessary for such a determination. In the current study, we used the computations, confirmed by experiments, from Alberti et al.\cite{alberti2019modeling} Our future experiments will involve a re-arrangement of our experimental platform to use our streak camera to obtain a more complete characterization of the plasma kernel shape variation with time.

\begin{acknowledgments}
We wish to acknowledge the support provided through the NSF/DOE  Partnership in Basic Plasma Science and Engineering, grant number DE-SC0020068, with Dr. Nirmol Podder as the Program Manager.

The data that support the findings of this study are available from the corresponding author upon reasonable request.
\end{acknowledgments}

\nocite{*}
\bibliography{aipsamp}

\end{document}